\documentstyle[11pt]{article}
\begin{document}
\begin{center}
\hfill ITEP/TH-28/96\\
\hfill hep-th/9607162\\
\vspace{0.3in}

{\large \bf BETHE ANSATZ AND CLASSICAL HIROTA EQUATIONS}
\footnote{Talk presented at the II Sakharov International Conference}
\end{center}

\centerline{\large \bf A.V.Zabrodin}

\begin{center}
{\it Joint Institute of chemical physics, Kosygina str. 4,
117334, Moscow, Russia\\
and\\
ITEP, B.Cheremushkinskaya 25, 117259, Moscow, Russia}
\end{center}

\begin{quotation}{A brief non-technical review of the recent
study \cite{KLWZ} of classical integrable structures in quantum
integrable systems is given. It is explained how to identify
the standard objects of quantum integrable systems (transfer matrices,
Baxter's $Q$-operators, etc) with elements of classical non-linear
integrable difference equations ($\tau$-functions, Baker-Akhiezer
functions, etc). The nested Bethe ansatz equations for $A_{k-1}$-type
models emerge as discrete time equations of motion for zeros of
classical $\tau$-functions and Baker-Akhiezer functions. The connection
with discrete time Ruijsenaars-Schneider system of particles is
discussed.}
\end{quotation}

At present it is argued that classical and quantum integrable models have
much deeper interrelations than any kind of a naive "classical limit".
In \cite{KLWZ}, a particular aspect of this phenomenon has been
analysed in detail: Bethe equations, which are usually considered as a
tool inherent to quantum integrability, arise naturally as a result
of solving entirely {\it classical} non-linear discrete time
integrable equations. This suggests an intriguing connection between
integrable quantum field theories and classical soliton equations
in discrete time.

Here we outline main ideas of the paper \cite{KLWZ} omitting all
technical details. To save the space, the material is organized
as a "quantum - classical dictionary" supplied with brief comments.

\bigskip

{\bf 1. Eigenvalues of quantum transfer matrices $=$ (classical)
$\tau$-functions}. Due to the Yang-Baxter equation the transfer matrices
commute for all values of the spectral parameters in the auxiliary
space (AS)\footnote{Here and below we use the notions and terminology
of the quantum inverse scattering method, see e.g.\,\cite{QISM}.}:
$\phantom{a}[T_{{\cal A}}(u),\, T_{{\cal A}'}(u')]=0$. This property
allows one to diagonalize them simultaneously. From now on we use this
diagonal representation. The identification with $\tau$-function
is justified in the next item.

\bigskip

{\bf 2. Fusion rules $=$ Hirota's difference equation}.
The fusion procedure in the AS gives rise to a family of commuting
transfer matrices $T_{{\cal A}}(u)$ with the same quantum space.
They obey a number of fusion relations \cite{GL3} which can be
recast into the model-independent bilinear form \cite{KP}.
Let $T_{s}^{a}(u)$ be the transfer matrix for the rectangular
Young diagram of length $s$ and height $a$, then it holds
\begin{equation}
T^{a}_{s}(u+1)T^{a}_{s}(u-1)-
T^{a}_{s+1}(u)T^{a}_{s-1}(u)=
T^{a+1}_{s}(u)T^{a-1}_{s}(u)\,.
\label{1}
\end{equation}
Remarkably, this equation coincides with Hirota's bilinear difference
equation (HBDE) \cite{Hirota} which is known to unify the majority
of soliton equations, both discrete and continuous.

Fusion of more complicated representations in the AS is described by
higher representatives of the hierarchy of HBDE-like equations
\cite{DJM}.
For example, consider Young diagrams consisting of two rectangular
blocks (i.e. with $a_1$ lines of length $s_1 +s_2$ and the rest
$a_2$ lines of length $s_1$) and let $T_{s_{1}, s_{2}}^{a_1 , a_2 }(u)$
be the corresponding transfer matrix. Then it holds
\begin{eqnarray}
&&T_{s_1 , s_2 }^{a_1 , a_2 -1}(u)
T_{s_1 -1, s_2 -1}^{a_1 , a_2 +1}(u)+
T_{s_1 , s_2 +1}^{a_1 -1, a_2 -1}(u)
T_{s_1 , s_2 -1}^{a_1 +1, a_2 +1}(u) \nonumber \\
&+&T_{s_1 +1, s_2 }^{a_1 -1, a_2 }(u-1)
T_{s_1 -1, s_2 }^{a_1 +1, a_2 }(u+1) \nonumber \\
&=&T_{s_1 +1, s_2 }^{a_1 , a_2 -1}(u-1)
T_{s_1 -1, s_2 }^{a_1 , a_2 +1}(u+1)+
T_{s_1 , s_2 +1}^{a_1 -1, a_2 }(u-1)
T_{s_1 , s_2 -1}^{a_1 +1, a_2 }(u+1)\,.
\label{2}
\end{eqnarray}
In what follows we consider eq.\,(\ref{1}) only.

\bigskip

{\bf 3. Specifying a particular quantum model $=$ imposing
particular boundary and analytic conditions in HBDE}.

For models associated to $A_{k-1}$-type quantum algebras these
conditions are:
\begin{eqnarray}
&&T^{a}_{s}(u)=0 \;\;\;\;\mbox{as}\;\;a<0 \;\;\mbox{or}\;\; a>k;
\nonumber \\
&&T_{s}^{a}(u)=0
\;\;\;\;\mbox{for any}\;\; -k<s<0,\;\;0<a<k\,.
\label{3}
\end{eqnarray}
It follows from (\ref{1}), (\ref{3}) that $T_{s}^{0}(u)$ and
$T_{s}^{k}(u)$ factorize into a product of functions of $u+s$ and
$u-s$. We adopt the normalization in which
$T_{s}^{0}(u)=\phi (u+s)$,
$T_{s}^{k}(u)=\phi (u-s-k)$, where the function $\phi (u)$ carries
information about the particular quantum model.

Another important condition
(which follows, eventually, from the Yang-Baxter
equation) is that $T^{a}_{s}(u)$ for models with elliptic
$R$-matrices has to be an
elliptic polynomial in the spectral parameter $u$. (By elliptic polynomial
we mean essentially a finite product of Weierstrass $\sigma$-functions.)
For models with rational $R$-matrix it degenerates to a usual
polynomial in $u$.

\bigskip

{\bf 4. "Step" of nested Bethe ansatz $=$ B\"acklund transformation}.
Quantum integrable models with internal degrees of freedom can be solved
by the nested (hierarchical) Bethe ansatz method. The method consists
essentially in integration over a part of degrees of freedom by an ansatz
of Bethe type, the effective hamiltonian being again integrable.
Repeating this step several times, one reduces the model to an integrable
model without internal degrees of freedom which is solved by the usual
Bethe ansatz.

The classical face of this scheme is B\"acklund transformation,
i.e. passing from solutions to the non-linear equation to (properly
normalized) solutions of the auxiliary linear problems (ALP), which
satisfy the same non-linear equation. The ALP for HBDE have the form
\cite{Hirota}
\begin{eqnarray}
&&T_{s+1}^{a+1}(u)F^{a}(s, u)-
T_{s}^{a+1}(u-1)F^{a}(s+1, u+1)=
T_{s}^{a}(u)F^{a+1}(s+1, u)\,, \nonumber\\
&&T_{s+1}^{a}(u-1)F^{a}(s, u)-
T_{s}^{a}(u)F^{a}(s+1, u-1)=
T_{s}^{a+1}(u-1)F^{a-1}(s+1, u)\,,
\label{4}
\end{eqnarray}
Eq.\,(\ref{1}) is the compatibility condition for (\ref{4}). Due to
the symmetry between $T$ and $F$ \cite{SS},
the latter satisfies {\it the
same} non-linear equation (\ref{1}),
so the transition $T\rightarrow F$ is a
B\"acklund transformation.
The important point is that the boundary condition
for $F^{a}(s,u)$ is the same as in (\ref{3}), the only change being a
reduction of the Dynkin graph: $k\rightarrow k-1$.  In other words, the
number of non-zero functions gets reduced by 1.
Using this property, one can
successively reduce the $A_{k-1}$-problem up to $A_1$.

\bigskip

{\bf 5. Eigenvalues of Baxter's $Q$-operators $=$ properly normalized
solutions to the ALP (Baker-Akhiezer functions)}.
To elaborate the chain of B\"acklund transformations, let
$t=0,\, 1,\ldots ,k$ mark steps of the flow
$A_{k-1}\rightarrow A_1$
and let
$F^{a}_{t+1}(s,u)$ be a solution to the ALP at $(k-t)$-th
step (in this notation $F_{k}^{a}(s,u)=T_{s}^{a}(u)$)
such that $F^{a}_{t}(s,u)=0$ as $a<0$ or
$a>t$. Due to this condition $F_{t}^{0}$ and $F_{t}^{t}$ are "chiral"
functions, i.e.
\begin{equation}
F_{t}^{0}(s,u)= Q_t (u+s)\,, \;\;\;\;\;
F_{t}^{t}(s,u)=  Q_t (u-s-t)\,,
\label{5}
\end{equation}
where $Q_t (u)$ are some functions playing a distinguished role in what
follows since they can be identified
with generalized Baxter's $Q$-operators
in the diagonal representation (see below).
Here is an example of this array
of $\tau$-functions for the $A_2$-case
($k=3$):

\vspace{2mm}

\begin{equation}
\begin{array}{ccccccccccc} 0&&1&&0&&&&&&\\ &&&&&&&&&&\\
0&&Q_1 (u+s)&& Q_1 (u-s-1)&&0&&&&\\
&&&&&&&&&&\\
0&&Q_2 (u+s)&&F_{2}^{1}(s,u)&&  Q_2 (u-s-2)&&0&&\\
&&&&&&&&&&\\
0&&Q_3 (u+s)&&F_{3}^{1}(s,u)&&F_{3}^{2}(s,u)&& Q_3 (u-s-3)&&0\\
&&&&&&&&&&\\
\end{array}
\label{diag}
\end{equation}
(in this case $Q_3 (u)=\phi (u)$).

\bigskip

{\bf 6. Nested Bethe ansatz equations $=$ Calogero-type models
in discrete time}. It follows from (\ref{4}) that
$\tau ^{t,a}(u)=F^{a}_{k-t}(u+a, u)$ satisfies the bilinear equation
\begin{equation}
\tau ^{t+1,a}(u) \tau ^{t, a+1}(u)-
\tau ^{t,a}(u) \tau ^{t+1, a+1}(u)=
\tau ^{t+1,a}(u+1) \tau ^{t, a+1}(u-1)
\label{6}
\end{equation}
which is HBDE in "light cone" variables. At the same time
$\tau ^{t,0}(u)=Q_{k-t}(2u)$. Let $u_{j}^{t}$ be zeros of $Q_t (u)$:
$Q_{t}(u_{j}^{t})=0$. For models with elliptic $R$-matrices
$Q_{t}(u)$ should be elliptic polynomials in $u$. This condition
leads to a number of constraints for $u_{j}^{t}$ which can be derived
using the technique developed in \cite{Kr} for elliptic solutions to
the KP equation. These constraints are nothing else than the
nested Bethe ansatz equations:
\begin{equation}
\frac
{Q_{t-1}(u_{j}^{t}+2)Q_t (u_{j}^{t}-2)Q_{t+1}(u_{j}^{t})}
{Q_{t-1}(u_{j}^{t})Q_t (u_{j}^{t}+2)Q_{t+1}(u_{j}^{t}-2)}=-1
\label{7}
\end{equation}
(with the boundary condition
$Q_{0}(u)=1$, $Q_k (u)=\phi (u)$).
These equations can be understood as
"equations of motions" for zeros
of $Q_{t}(u)$ in discrete time $t$ (level of the Bethe ansatz
which runs over the Dynkin graph). The analogy with elliptic
solutions of the KP equation suggests to call them the discrete time
analogue of the Ruijsenaars-Schneider (RS) system of particles (see also
\cite{NRK}). Taking the continuum limit in $t$ (provided the number
of zeros $M_t =M$ of $Q_{t}(u)$ in a fundamental domain does not
depend on $t$), one can verify that this system of equations does
yield equations of motion for the RS system \cite{RS} with $M$
particles.

However, integrable systems of particles in discrete time have a
richer structure than their continuous counterparts. In particular,
the total number of particles may depend on the discrete time. Such a
phenomenon is possible in continuous time models only for singular
solutions, when particles can move to infinity or merge to another
within a finite period of time. Remarkably, this appears to be the
case for the solutions to eq.\,(\ref{7}) corresponding to
eigenstates of quantum models. It is known that the number of
excitations at $t$-th level of the nested Bethe ansatz solution
does depend on $t$. In other words, the number of "particles" in
the associated discrete time RS system is not conserved. At the
same time the numbers $M_t$ may not be arbitrary. It can be shown
that for models with elliptic $R$-matrices in case of general
position $M_t = (N/k)t$, where $N$ is the number of sites of the
lattice (degree of the elliptic polynomial $\phi (u)$). In trigonometric
and rational cases the conditions on $M_t$ become less restrictive
but still they may not be equal to each other.

\vspace{2mm}

At last we should indicate how to identify our $Q_t$'s with
$Q_t$'s from the usual nested Bethe ansatz solution. This is
achieved by the following factorization formula which allows
one to express $T_{s}^{a}(u)$ in terms of $Q_t$'s:
\begin{eqnarray}
&&\sum_{a=0}^{k}(-1)^{a-k}
\frac{T^{a}_{1}(u+a-1)}{\phi (u-2)}e^{2a\partial _{u}}=\nonumber \\
&\!\!\!=\!\!\!&\!\!\!\left ( e^{2\partial _{u}}-
\frac{Q_k (u)Q_{k-1}(u-2)}
{Q_k (u-2)Q_{k-1}(u)}\right )\! \ldots   \!
\left (e^{2\partial _{u}}-
\frac{Q_{2}(u)Q_{1}(u-2)}{Q_{2}(u-2)Q_{1}(u)} \right )\!
\left (e^{2\partial _{u}}-
\frac{Q_1(u)}{Q_1(u-2)}
\right ),
\label{factor}
\end{eqnarray}
where the shift operator
$e^{\partial _u}$
acts as usual:
$e^{\partial _u}f(u)=f(u+1)$. For the proof see ref.\,\cite{KLWZ}.
This formula coincides with the one known in the literature
(see e.g. \cite{BR1}). The left hand side is known as the generating
function for $T_{1}^{a}(u)$.

\vspace{2mm}

There is no doubt that this dictionary can be continued to include
not only spectral properties of quantum system on finite lattices
(which just correspond to elliptic solutions of HBDE) but dynamical
properties and correlation functions as well. Perhaps this amounts
to considering solutions to the same Hirota equation of a more
complicated type. Another intriguing problem is to extend this
dictionary to off-diagonal elements of quantum monodromy matrices
and quantum $L$-operators and identify them with some objects
in hierarchies of classical integrable equations.

\subsection*{Acknowledgements}

The author is grateful to I.Krichever,
O.Lipan and P.Wiegmann for collaboration in \cite{KLWZ} and to
B.Enriquez, A.Gorsky, S.Khoroshkin, A.Marshakov,
A.Mironov, N.Reshetikhin, N.Slavnov,
T.Takebe for discussions. This work was supported in part by
RFFR grant 96-01-01106, by ISTC grant 015, by INTAS 94 2317 and
the grant of the Dutch NWO organization.


\begin{thebibliography}{10}

\bibitem{KLWZ} I.Krichever, O.Lipan, P.Wiegmann and A.Zabrodin,
preprint ESI 330 (1996), hep-th/9604080.

\bibitem{QISM}
L.D.Faddeev and L.A.Takhtadjan, Uspekhi Mat. Nauk {\bf 34:5} (1979) 13;
P.P.Kulish and E.K.Sklyanin, Lecture Notes in Physics {\bf 151}
61, Springer, 1982.

\bibitem{Faddeev}
L.D.Faddeev and L.A.Takhtadjan, {\it Quantum inverse scattering method
and the $XYZ$ Heisenberg model}, Uspekhi Mat. Nauk {\bf 34:5} (1979) 13-63.


\bibitem{GL3} P.P.Kulish and N.Yu.Reshetikhin,
Zap. Nauchn. Sem. LOMI {\bf 120}
(1982) 92 (in russian).

\bibitem{KP} A.Kl\"umper and P.Pearce,
Physica
{\bf A183} (1992) 304;
A.Kuniba, T.Nakanishi and J.Suzuki,
Int. Journ. Mod.
Phys. {\bf A9} (1994) 5215.

\bibitem{Hirota} R.Hirota, Journ. Phys. Soc. Japan {\bf 50} (1981) 3785.

\bibitem{DJM} E.Date, M.Jimbo and T.Miwa, Journ. Phys. Soc. Japan
{\bf 53} (1982) 4116.

\bibitem{SS} S.Saito and N.Saitoh, Phys. Lett.
{\bf A120} (1987) 322.

\bibitem{Kr}
I.M.Krichever, Func. Anal. App {\bf 14}
(1980), n 4, 282.

\bibitem{NRK}
F.Nijhof, O.Ragnisco and V.Kuznetsov, Commun. Math. Phys. {\bf 176} (1996)
681.

\bibitem{RS} S.N.M.Ruijsenaars and H.Schneider, Ann. Phys. (NY)
{\bf 170} (1996) 681.

\bibitem{BR1} V.Bazhanov and N.Reshetikhin, Journ. Phys. {\bf A23}
(1990) 1477.

\end{thebibliography}
\end{document}